\documentclass[runningheads]{llncs}

\usepackage[T1]{fontenc}
\usepackage{graphicx}
\usepackage{hyperref}
\usepackage{color}

\usepackage{array}
\usepackage{multirow}
\usepackage{amsmath}
\usepackage{amsfonts}

\makeatletter
\renewcommand{\@fnsymbol}[1]{\ifcase#1\or*\fi}
\makeatother

\begin{document}


\title{Cine cardiac MRI reconstruction using a convolutional recurrent network with refinement} 

\titlerunning{Reconstruction of single coil multi-view cine cardiac MRI}

\author{Yuyang Xue\inst{1}\thanks{These authors contributed equally.} \and Yuning Du\inst{1}\textsuperscript{*} \and Gianluca Carloni\inst{2,3}\textsuperscript{*} \and Eva Pachetti\inst{2,3}\textsuperscript{*} \and Connor Jordan\inst{4}\textsuperscript{*} \and Sotirios A. Tsaftaris\inst{1,5}}


\authorrunning{Y. Xue et al.}

\institute{Institute for Digital Communications, University of Edinburgh, United Kingdom\\
\and Institute of Information Science and Technologies ``Alessandro Faedo'', National Research Council of Italy (CNR), Pisa, Italy\\
\and Department of Information Engineering, University of Pisa, Italy\\
\and Institute for Infrastructure and Environment, University of Edinburgh, United Kingdom\\
\and The Alan Turing Institute, London, United Kingdom}

\maketitle              


\begin{abstract}
Cine Magnetic Resonance Imaging (MRI) allows for understanding of the heart's function and condition in a non-invasive manner. 
Undersampling of the $k$-space is employed to reduce the scan duration, thus increasing patient comfort and reducing the risk of motion artefacts, at the cost of reduced image quality.
In this challenge paper, we investigate the use of a convolutional recurrent neural network (CRNN) architecture to exploit temporal correlations in supervised cine cardiac MRI reconstruction. 
This is combined with a single-image super-resolution refinement module to improve single coil reconstruction by 4.4\% in structural similarity and 3.9\% in normalised mean square error compared to a plain CRNN implementation.
We deploy a high-pass filter to our $\ell_1$ loss to allow greater emphasis on high-frequency details which are missing in the original data.
The proposed model demonstrates considerable enhancements compared to the baseline case
and holds promising potential for further improving cardiac MRI reconstruction.

\end{abstract}

\section{Introduction}
\label{sect:intro}
Cardiac magnetic resonance imaging is a powerful, non-invasive tool to aid visualise the heart's chambers, valves, blood vessels and surrounding tissue.
To gain a 3D depiction of the heart, a sequential acquisition process of 2D slices is used, with the scanning duration increasing with the number of slices and temporal resolution desired.
Thus for detailed scanning, multiple cardiac cycles must be monitored and the duration of the MRI process can consequently exceed the patients ability to remain steady and hold their breath.
By undersampling in the $k$-space data acquisition process, the scan time can be substantially reduced at the cost of missing information that must be interpolated.
Deep learning achieves $k$-space reconstruction with greater prior knowledge for the regularisation term that covers the missing $k$-space domain, without requiring an iterative optimisation process and hence greatly accelerating the reconstruction rate. 

Various architectures have been explored for MRI reconstruction, including convolutional neural networks (CNNs) and U-Nets \cite{Han2018,Hyun2018,Lee2018,Wang2016}, variational networks \cite{Hammernik2018} and generative adversarial networks \cite{Lyu2023,Yang2018}.
Other deep learning methods that exploit prior knowledge and extend traditional iterative methods include the model-based deep-learning architecture \cite{Aggarwal2019} and deep density priors \cite{Tezcan2019}.
Enhancing cine MRI through deep learning involves not only capitalising on the spatial relationships acquired from a given dataset but also leveraging temporal correlations.
This has been evidenced across various model architectures \cite{Kofler2021,Lyu2023,Vornehm2022,Zhang2022} as well as through registration-based \cite{Yang2022} and motion-guided alignment \cite{Han2023} approaches.
In \cite{Schlemper2018}, data sharing layers were incorporated in a cascaded CNN, whereby adjacent time step $k$-space data was used to fill the unsampled lines.
In \cite{Qin2019}, recurrent connections are employed across each iteration step as well as bidirectional convolutional recurrent units facilitating knowledge sharing between iterations and input time frames, respectively.

Working within the confines of the challenge, we explored various architectures and found the CRNN block of \cite{Qin2019} to perform best within the given limitations in memory and reconstruction time set by the organisers.
This was subsequently combined with a lightweight refinement module inspired by single-image super-resolution approaches \cite{Bilecen2023} to perform further de-noising and resolve finer details.
The rest of the paper is organised as follows: Sections \ref{S2:prob} and \ref{S3:mthd} describes the problem, dataset, and methodology, Sect. \ref{S4:res} presents the results of experiments and Sect. \ref{S5:disc} and \ref{S6:cncl} provide discussion and conclusions, respectively.

\section{Problem formulation and dataset}
\label{S2:prob}
The objective of MRI reconstruction is to address an ill-posed inverse problem, retrieving image information denoted as $\mathbf{x} \in \mathbb{C}^{N}$ from acquired undersampled signals $\mathbf{y} \in \mathbb{C}^{K}$, where $K \ll N$. 
This procedure can be depicted using a linear forward operator $\mathbf{E}$, which defines the characteristics of the forward problem:
\begin{equation} \label{eq1}
    \mathbf{y} 
    = \mathbf{E} \mathbf{x} + \epsilon .
\end{equation} 
Eq.~\ref{eq1} represents the general form of MRI reconstruction.
The goal of reconstruction is to minimise the difference between $\mathbf{x}$ and the ground truth. 
Therefore, the reconstruction problem can be defined as follows:

\begin{equation}
    \tilde{\mathbf{x}} = \underset{\mathbf{x}\in \mathbb{C}^{N}}{\arg \min}  \frac{\lambda}{2}\|\mathbf{Ex}-\mathbf{y}\| + f_\theta(\mathbf{x}).
\end{equation}

Here, $f_\theta$ denotes a neural network for image reconstruction with trainable parameters $\theta$ and $\lambda$ controls the balance between the network and data consistency.

\subsubsection{Data}
Our model is evaluated on the CMRxRecon Challenge dataset from the 26\textsuperscript{th} International Conference on Medical Image Computing and Computer Assisted Intervention. 
The dataset includes both short-axis (SA) and long-axis (LA) (two-chamber, three-chamber and four-chamber) views under acceleration rates of $4\times$, $8\times$, $10\times$.
The dataset was obtained following recommended protocols and processing \cite{Wang2021,Zhang2013}, more details of which can be found on the challenge website \cite{CMRxRecon2023}.
The 300-patient dataset is split 120:60:120 between challenge training, validation, and testing respectively.
Only the challenge training set contained ground truth reference data, hence this was further split 90:20:10 for training, evaluation, and testing respectively for all models.

\subsubsection{Data pre-processing}
The unpadded image size varies between widths of 132, 162, 204 \& 246 and heights of 448 \& 512 pixels.
To maintain consistent input size, we apply zero-padding for image sizes of 256 $\times$ 512 following the Inverse Fourier Fast Transform, with the outputs cropped after inference.

When using the approach from \cite{Kofler2021}, the computationally intensive conjugate gradient step was a limiting factor due to the GPU's initial 24GB memory constraint set by the challenge organisers.
We thus chose to use the coil combined data rather than the multi-coil format, which allowed use of a simpler data consistency step, at a potential loss of accuracy without using the extra information.
Likewise, we used a single channel for the processed image instead of using independent channels for amplitude/phase or real/imaginary components, as adopted by \cite{Hammernik2018,Han2018,Lee2018,Schlemper2018}.

The SA data are 3-dimensional spatially with an additional time component.
It is therefore conceivable that full 4D convolutional kernels could be used to fully utilise spatio-temporal redundancies, but this would result in extremely large memory requirements as discussed in \cite{Kustner2020}.
Furthermore, studies such as \cite{Vornehm2022} have demonstrated that it is preferable to have a larger $2\mbox{D} + t$ network than a smaller 3D-input network with equivalent memory consumption.
Therefore, due to the large image size, we choose to use time-series batches of 2D depth slices as per \cite{Qin2019} rather than $3\mbox{D} + t$ or 4D for the long-axis images.

\section{Methodology}
\label{S3:mthd}

\subsection{Model exploration}
The initial limitations for inference imposed by the organisers were 24GB GPU VRAM and 4 hours for the reconstruction of the test dataset, which was later increased after our initial investigations.
Pre-trained models or loss functions were not permitted. 
Denoising diffusion probabilistic models (DDPM) were found to take too long in inference, whilst transformer models have been found to lead to heavily pixelated reconstructions.
Hence, more conventional approaches were tested, building upon an existing repository\footnote{\url{https://github.com/f78bono/deep-cine-cardiac-mri}}.

We compared networks similar to the CineNet \cite{Kofler2021} and CRNN networks \cite{Qin2019}.
The number of parameters in each network were maximised such that the full 24GB VRAM memory would be used in training.
A 2D U-Net is deployed to serve as an additional baseline to compare all models to.
The U-Net is trained on a slice-by-slice basis with 3 cascades and 48 feature map channels. 
Weight sharing is used when training the model, and $\lambda$ is set to be learnable with an initialisation of $\mathrm{log}(10^{-1})$. 
The learning rate is $3\times 10^{-4}$, and the Adam optimiser is deployed to guide the training process.

\vspace{-2mm}

\subsection{Model architecture}
\begin{figure}[!b]
    \centering
    \includegraphics[width=\textwidth]{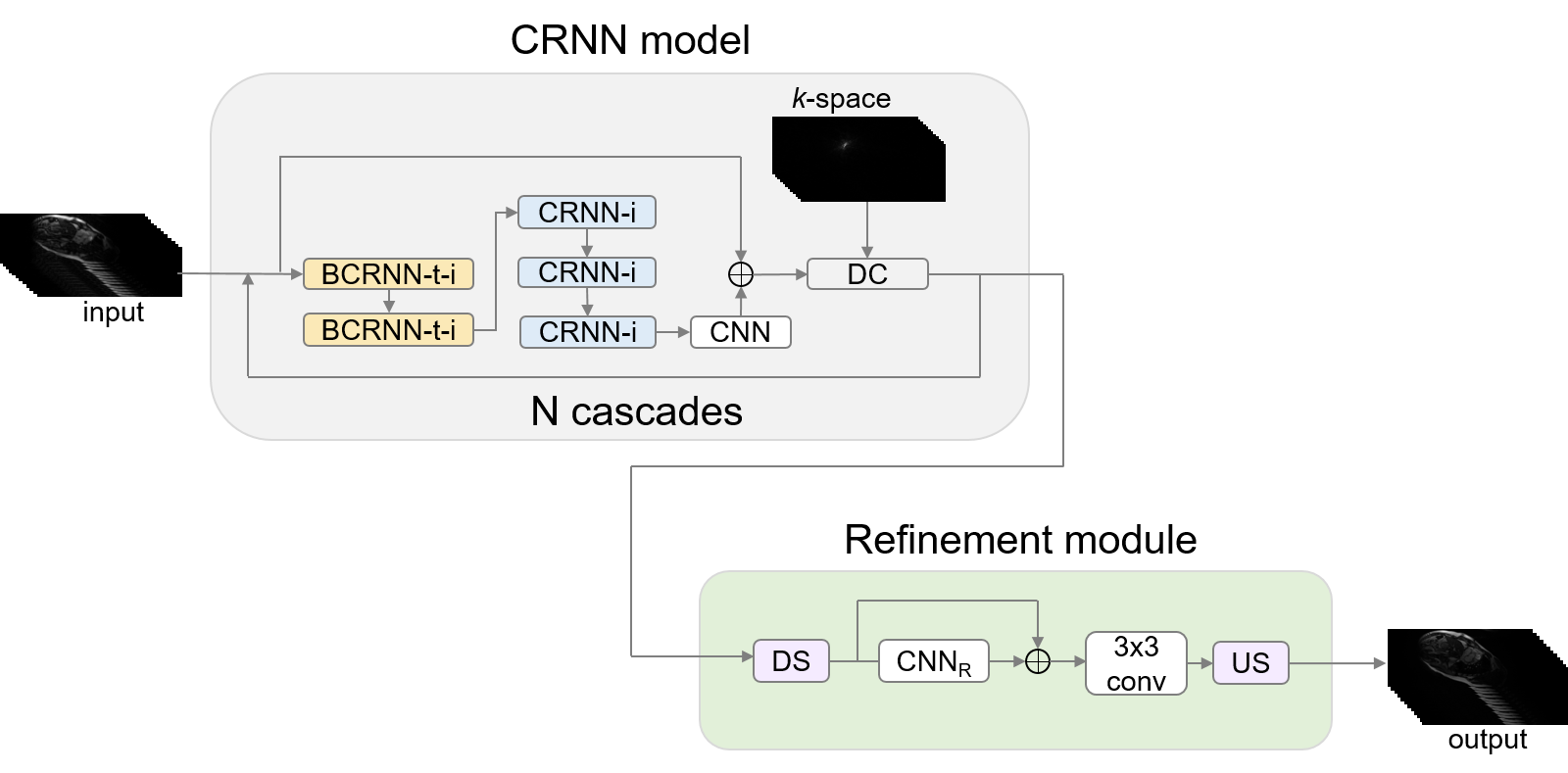}
    \setlength{\abovecaptionskip}{-10pt}
    \caption{Final model architecture: BCRNN, CRNN, and CNN units with a data consistency (DC) step from \cite{Qin2019} for primary reconstruction. "t" and "i" denote time and iterations, respectively. The low-cost refinement module, inspired by \cite{Bilecen2023}, includes downsampling (DS), CNN, and upsampling (US) units.}
    \label{fig:model_arch}
\end{figure}

A high-level depiction of the complete architecture of the final model is presented in Fig. \ref{fig:model_arch}.
The backbone of proposed architecture is based on the CRNN block detailed in \cite{Qin2019}. 
The first step in CRNN is a bidirectional convolutional recurrent unit (BCRNN) with three convolution layers: a standard convolution between layers, one convolution between temporal slices, and one between iterations.
This is followed by three convolutional recurrent units (CRNN) which evolve only over iterations before a plain CNN.
Finally, residual connections are employed prior to a data consistency term, preserving the information from sampled data.

Aiming to improve performance, we include an additional BCRNN unit to further exploit spatio-temporal dependencies, followed by a refinement module to further denoise the output of the CRNN model and refine further details.
We deploy a very lightweight single-image super-resolution network, Bicubic++ \cite{Bilecen2023}, which maintains short reconstruction times.
The refinement module first learns lower resolution features to decrease computational cost and then performs numerous convolutions to denoise the image before a final convolutional filter and upscaling back to the original image size.
We test the performance of the network with end-to-end and separate learning for each module.

\vspace{-2mm}

\subsubsection{Loss function}
\label{SSS:loss}
In the context of of image reconstruction, $\ell_1$ loss, $\ell_2$ loss and SSIM loss are widely used to constrain models for high-quality reconstructed images, but often disregard the complex nature of MRI data. 
Thus, we investigate a range of losses using an additional loss term, denoted the $\perp$-loss \cite{Terpstra2022}.
The $\perp$-loss adds a phase term which can be combined with $\ell_1$, $\ell_2$ or SSIM losses to address the asymmetry in the magnitude/phase loss landscape.
This operates on the polar representation of complex numbers, rather than on two real value channels for magnitude and phase, thus taking advantage of the fact that fully symmetric loss functions can improve task performance \cite{Patel2021}.
For the separate training of the CRNN and the refinement module, $\perp$-losses are only utilised for the CRNN output, with $\ell_1$ and SSIM loss functions deployed for the refinement module.
For the end-to-end training, $\ell_1$ and SSIM loss are employed to constrain both CRNN and refinement module.
We split the $\ell_1$ loss by introducing a high-pass frequency filter, allowing us to emphasise the high-frequency content in our reconstructed images to resolve finer details.
We denote this as $\ell_1$\textsubscript{split}.

In training, losses were quantified across the entire image, whereas for the validation leaderboard, assessment was limited to the initial 3 time frames and the central sixth portion of the images.
The competition metrics were structural similarity index measure (SSIM), normalised mean square error (NMSE) and peak signal-to-noise ratio (PSNR). 
Hence, whilst the complete reconstructed images often surpassed SSIM values of 0.98, validation scores only reached 0.85.
\vspace{-4mm}

\subsection{Implementation details}
Implementation details and our code is available at: \url{https://github.com/vios-s/CMRxRECON_Challenge_EDIPO}

\vspace{-1mm}

\section{Results}
\label{S4:res}

\subsubsection{Model choice and weight sharing}
\begin{figure}[!t]
    \centering
    \includegraphics[width=\textwidth]{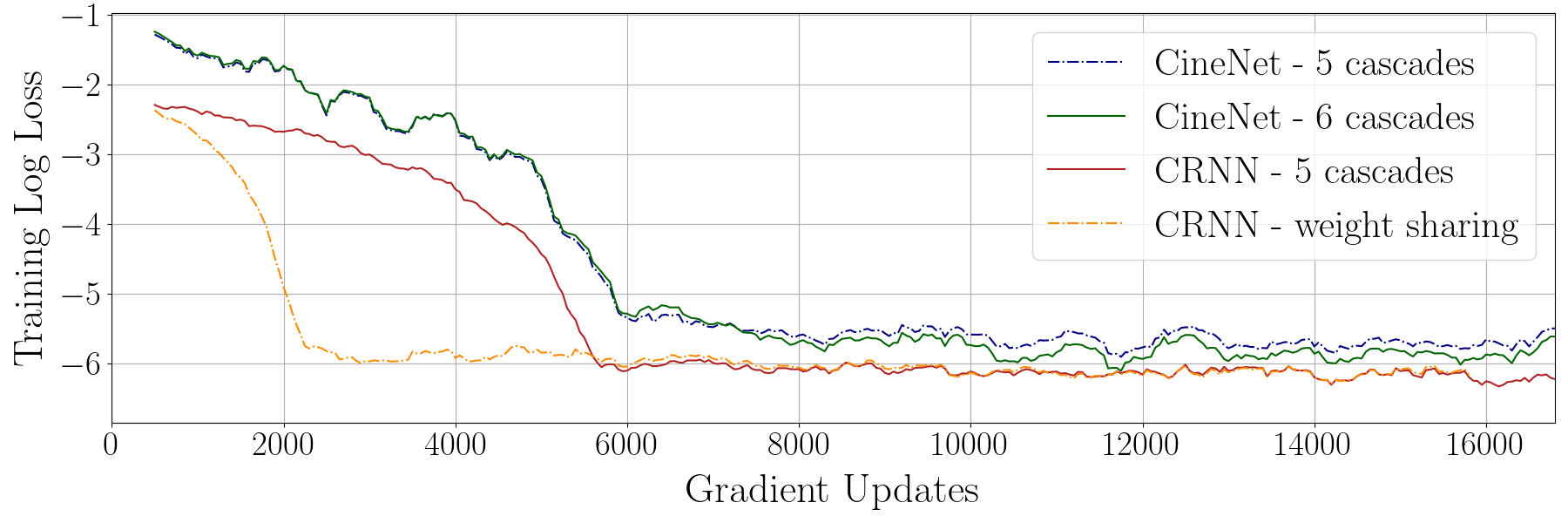}
    \setlength{\abovecaptionskip}{-15pt}
    \caption{Log loss during exploratory training of modified CineNet and CRNN (with and without weight-sharing between kernels). Note that the implementation is not identical to the original works.}
    \label{fig:traininglosses}
\end{figure}

\begin{figure}[!t]
    \centering
    \includegraphics[width=\textwidth]{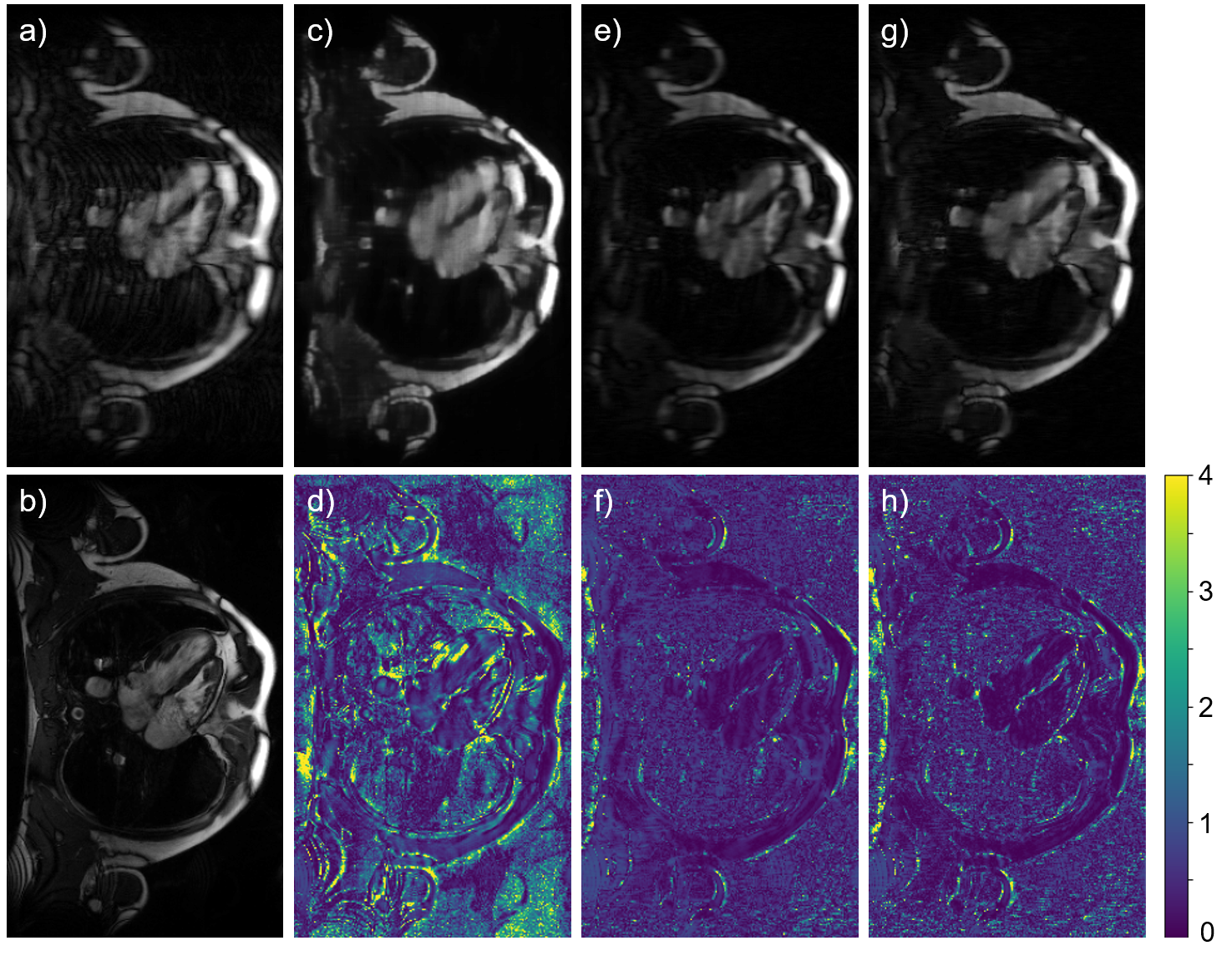}
    \setlength{\abovecaptionskip}{-15pt}
    \caption{Reconstruction (top) and associated error maps (bottom) for the initial network investigation. (a) 8 $\times$ undersampled LAX input (b) fully sampled ground truth (c,d) CineNet model (6 cascades) (e,f) CRNN model (weight-sharing between cascades) (g,h) CRNN model (no weight-sharing).}
    \label{fig:visualisation_initial}
    \vspace{-5mm}
\end{figure}

Fig. \ref{fig:traininglosses} shows the training losses between models, demonstrating the stronger performance of the CRNN model.
The point of convergence for the CRNN with weight-sharing and the CineNet models is similar, but the CRNN networks start at a much lower loss value.
This is despite the non-weight-sharing model (1.1M) having over 2$\times$ more trainable parameters than the 6-cascade CineNet model (0.5M). 
Between CRNN models, we see more rapid convergence in the weight-sharing model as there are less parameters to optimise and reduced likelihood of early overfitting.
However, the lower number of parameters leads to reduced expressive power and is outperformed by the non-weight-sharing model with insufficient memory gains to justify its use.
Using the $\perp$-loss only, the weight-sharing model had SSIM of 0.683, NMSE of 0.123 and PSNR of 23.917, performing notably worse than the non-weight-sharing model, as presented in the next section.
Fig. \ref{fig:visualisation_initial} shows the reconstruction through the CineNet and the CRNN (with and without weight-sharing) models. 

\subsubsection{Loss function investigation}

Table \ref{tab:lossfunction} presents the findings of investigations of the loss functions using a low-cost CRNN model. 
Use of the $\perp$-$\ell_1$ loss led to an improvement in SSIM and PSNR compared to $\ell_1$ loss alone, but a slightly higher NMSE error.
Providing greater emphasis on high-frequency data using a high-pass filter led to improved SSIM but lower NMSE and PSNR performance.
Further tuning to improve the ratio of high- to low- frequency led to better results for the higher cascade models.
Notably, combining SSIM with the $\perp$-$\ell_1$ loss was counter-productive for all metrics and suggests that further tuning of the weighting of each loss component is required.

\begin{table*}[!tb]
\centering
\caption{Performance comparisons for different loss function combinations, evaluated on the validation data. A 48-channel 5 cascade CRNN network was used without the refinement module. $\ell_1$\textsubscript{split} denotes the $\ell_1$ loss whereby a high-pass filter is used to provide more focus on the high frequency content, in addition to the conventional $\ell_1$ loss.}
\begin{tabular}{c @{\hspace{5pt}} c @{\hspace{10pt}} c @{\hspace{10pt}} c @{\hspace{10pt}} c @{\hspace{10pt}} c @{\hspace{10pt}}}
\hline
 & & & & & \\ [-8pt]
\multirow{2}{*}{Metric} & \multicolumn{5}{c}{Loss function} \\
  \cline{2-6}
& $\perp$ & $\ell_1$ & $\perp$-$\ell_1$ & $\perp$-$\ell_1$\textsubscript{split} & $\perp$-SSIM-$\ell_1$\textsubscript{split} \\ [1.5pt]
\hline
& & & & & \\ [-6pt]
SSIM  & 0.712 & 0.741 & 0.752 & \textbf{0.753} & 0.739 \\
NMSE & 0.0925 & \textbf{0.0646} & 0.0655 &  0.0671 & 0.0719 \\ 
PSNR & 25.143 & 26.525  & \textbf{26.535} & 26.487 & 26.067 \\
\hline
\end{tabular}
\label{tab:lossfunction}
\end{table*}
\vspace{-2mm}
\subsubsection{Introduction of the refinement module}

Table \ref{tab:superresolution} highlights the improvements in the quality of inference made by introducing the refinement module. 
Deploying the refinement as a separately trained post-processing module shows notable benefit improving more than adding an additional cascade to the plain CRNN.
The end-to-end model results in further improvements upon separate training, of 4.4\% in structural similarity and 3.9\% in normalised mean square error relative to the plain CRNN, in spite of no longer being able to take advantage of the $\perp$ loss.
\vspace{-2mm}

\begin{table*}[!h]
\centering
\caption{Performance comparisons of various model set-ups. Sequential (separate) and end-to-end training (combined) of the CRNN and refinement module are presented.}
\begin{tabular}{c @{\hspace{10pt}} | @{\hspace{5pt}} c @{\hspace{10pt}} c @{\hspace{10pt}} c @{\hspace{10pt}} c @{\hspace{10pt}} c}
\hline
& & & &  \\ [-8pt]
Cascades & 6 & 6 & 6 & 7 \\
Refinement module & None & Sequential & End-to-end & None \\[1.5pt] 
\hline
& & & &  \\ [-6pt]
SSIM & 0.768 & 0.792 & \textbf{0.802} & 0.765 \\
NMSE & 0.0516 & 0.0496 & \textbf{0.0454} & 0.0535 \\ 
PSNR & 27.354 & 27.597 & \textbf{27.969} & 27.351 \\ [0.5ex]
\hline
\end{tabular}
\label{tab:superresolution}
\vspace{-2mm}
\end{table*}

\begin{figure}[!ht]
    \centering
    \includegraphics[width=\textwidth]{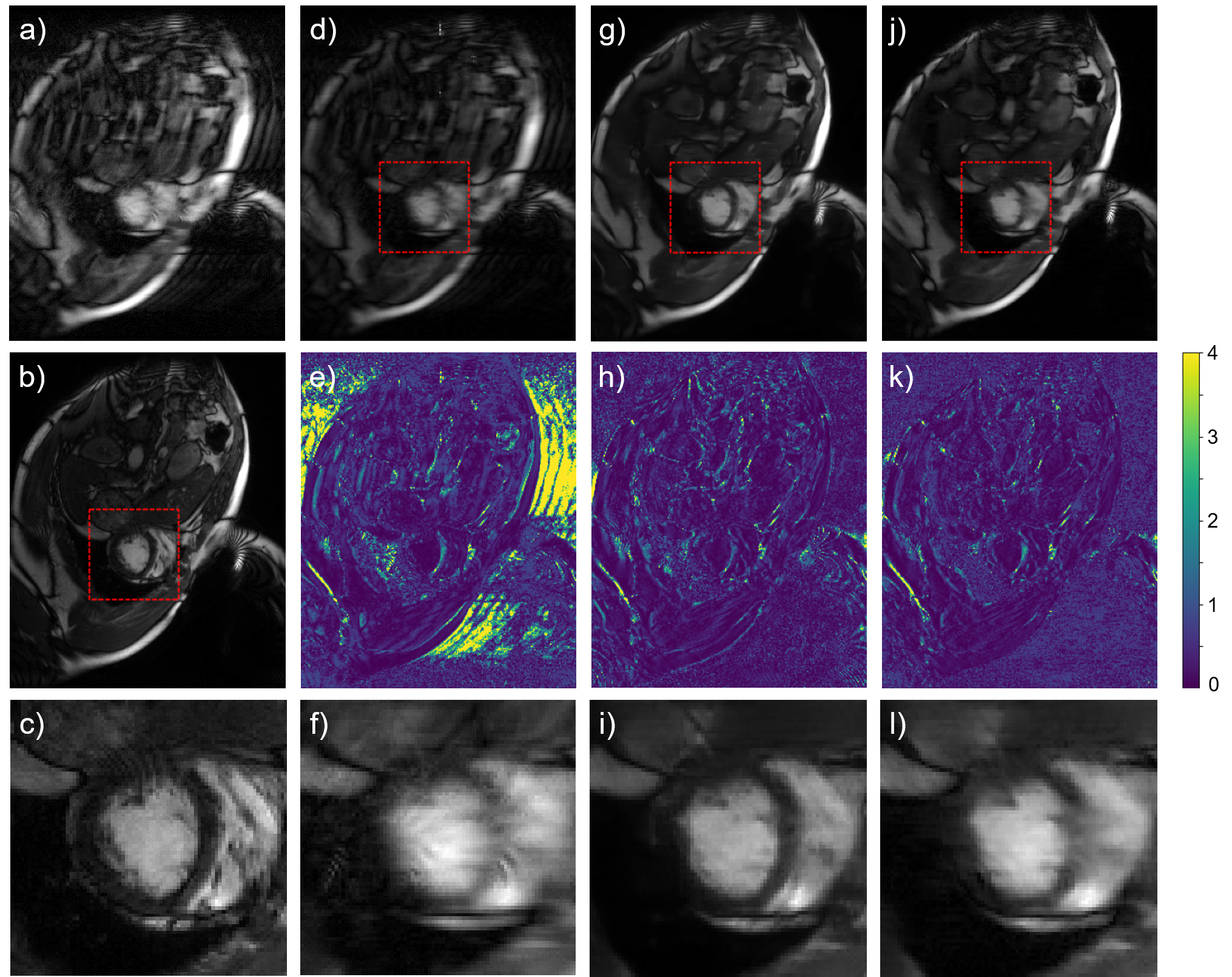}
    \setlength{\abovecaptionskip}{-4pt}
    \caption{Reconstruction (top) and associated error maps (middle) for the U-Net baseline and CRNN models. Finer details (bottom) are not resolved by the U-Net, which are partially captured by the plain CRNN model. The refinement module subsequently deblurs the image and provides better resolution at boundaries. (a) 10 $\times$ undersampled SAX input (b, c) fully sampled ground truth (d, e, f) U-Net (g, h, i) 6 cascades with combined refinement (j, k, l) 7 cascades, no refinement.}
    \label{fig:vis}
\end{figure}

Fig. \ref{fig:vis} shows qualitatively the improvements made by the introduction of the refinement module at full scale.
The error is reduced substantially and some finer details are resolved, though there is still scope for improvement at smaller scales.
We generally see that the model is incapable of generating details that are completely lost in the undersampling process.

\vspace{-2mm}
\subsubsection{Validation results}

Our final tests prior to submission are presented in Table \ref{tab:finalresults}.
Across all models, the short-axis reconstruction performs better quantitatively as there is more short-axis data available in training.
For both views, the performance reduces with increased undersampling, as more detail is lost.

\begin{table*}[!t]
\centering
\caption{Performance comparisons on evaluation of CMRxRecon cine cardiac MRI coil combined validation data for different acceleration rates (AR).}
\begin{tabular}{c @{\hspace{5pt}} c @{\hspace{10pt}} c @{\hspace{10pt}} c @{\hspace{10pt}} c @{\hspace{10pt}} c @{\hspace{10pt}} c @{\hspace{10pt}} c}
\multirow{3}{*}{AR} & \multirow{3}{*}{Metric} & \multicolumn{3}{c}{Short-axis} & \multicolumn{3}{c}{Long-axis} \\
&  & U-Net & Plain & \multirow{2}{*}{Proposed} & U-Net & Plain & \multirow{2}{*}{Proposed} \\ 
&  & Baseline & CRNN & & Baseline & CRNN & \\ 
\hline
& SSIM & 0.641 & 0.824 & \textbf{0.854} & 0.573 & 0.757 & \textbf{0.792} \\
$4\times$ & NMSE & 0.180 & 0.0311 & \textbf{0.0277} & 0.188 & 0.0485 & \textbf{0.0433} \\
& PSNR & 23.084 & 29.842 & \textbf{30.295} & 22.040 & 26.958 & \textbf{27.540} \\ 
\hline
& SSIM & 0.637 & 0.796 & \textbf{0.829} & 0.574 & 0.723 & \textbf{0.763}  \\
$8\times$ & NMSE & 0.201 & 0.0428 & \textbf{0.0377} & 0.191 & 0.0687 & \textbf{0.0586} \\
& PSNR & 22.603 & 28.364 & \textbf{29.002} & 22.234 & 25.588 & \textbf{26.370} \\ 
\hline
& SSIM & 0.641 & 0.788 & \textbf{0.822} & 0.588 & 0.717 & \textbf{0.753} \\
$10\times$ & NMSE & 0.210 & 0.0464 & \textbf{0.0408} & 0.198 & 0.0724 & \textbf{0.064} \\
& PSNR & 22.429 & 28.030 & \textbf{28.644} & 22.065 & 25.343 
 & \textbf{25.965} \\
\end{tabular}
\label{tab:finalresults}
\end{table*}

\newpage

\section{Discussion}
\label{S5:disc}

\subsubsection{On model choice} 
Without extensive hyperparameterisation, the CRNN architecture demonstrated more promising performance than the CineNet, both qualitatively and quantitatively as shown in Fig. \ref{fig:traininglosses} and \ref{fig:visualisation_initial}.
In this network, the temporal average is subtracted from each slice and then the residuals are transformed into $x-t$ and $y-t$ axes being passed through a 2D U-Net structure.
The 2D U-Net is computationally lightweight as the CineNet was originally designed for multi-coil radial acquisition processes, but even with an increased number of cascades fails to resolve finer details as clearly shown in Fig. \ref{fig:visualisation_initial}.
The recurrent connections exploit the temporal dependencies between slices more effectively than by attempting to capture these relationships through transformation and sparsification.

However, training of the plain CRNN has still resulted in lower perceptual quality than presented in the original implementation, which may have been further improved with more hyperparameterisation tuning.
The large image size of up to 246 $\times$ 512 proved challenging, and in our implementation limited us to 5 cascades of 48 channels for the original 24GB inference limitation for the CRNN network. 
In the original work \cite{Qin2019}, the model consisted of 10 cascades with 64 feature channels, which operated on smaller data with a smaller GPU.

\subsubsection{On the final model}
The plain CRNN implementation substantially outperforms the baseline 2D U-Net, which has over double the number of trainable parameters, demonstrating the importance of exploiting temporal correlations.
The reconstruction of the plain CRNN is a considerable improvement upon the 10$\times$ undersampled input, however smaller scale details that are resolved are blurry (e.g. Fig. \ref{fig:vis}l).
The introduction of the low-cost refinement module led to better results with further denoising as presented in Fig. \ref{fig:vis}i.
This shows promise for the implementation of lightweight single-image super-resolution models to assist in improving cardiac cine MRI reconstruction, either combined with the main reconstruction model or as a post-processing step.
Relative to the ground truth, we still see that finer details are being missed that have been obscured due to the undersampling process.
In \cite{antun2020instabilities}, numerous MRI reconstruction experiments were conducted to test current deep learning reconstruction models.
They proposed that ``networks must be retrained on any subsampling patterns'' for end to end CNN networks. 
In our approach, we trained all the acceleration rate together to get a stable but averaged reconstruction in a single model, subsequently resolving less finer details.
The failure in generating details that have been lost, could be better tackled by a generative model \cite{jalal2021robust}.
As such, performance on patient volumes where more aliasing artefacts were present was poorer.

Therefore, to improve the proposed model, there is potential that training exclusively for each view can improve the final details.
However, our model performs relatively well on the validation stage leaderboard for high acceleration factors, where finer details are more difficult to resolve, whilst we generally perform worse at lower acceleration factors. 
This suggests that whilst our model failed to generate some finer details, other architectures also struggled once these details were lost or heavily obscured.
There are numerous further modules that could have been implemented, had more time been available.

\subsubsection{On the loss function}
We found that introduction of the $\perp$ loss to the $\ell_1$ loss improved both SSIM and PSNR, though at the expense of a slightly reduced $\ell_1$ value itself.
Treating the weightings of the loss functions as learnable parameters could lead to improved results in all metrics, as anticipated due to the results presented in \cite{Terpstra2022}.
Likewise, the introduction of the high-pass filter loss to focus the $\ell_1$ loss on higher-frequency information increases the complexity of optimisation but was beneficial after the weightings were improved, though not presented quantitatively here.

\section{Conclusions}
\label{S6:cncl}

In this challenge, we deployed a CRNN network combined with a refinement module to perform MRI reconstruction of cardiac cine data.
We train the model for a range of acceleration factors and views using a high-pass filter to focus our loss on high-frequency details.
From the quantitative analysis of the evaluation data and from direct viewing of the validation portion of the training data, the refinement module provides additional image quality with improvements of around 4\% in all metrics relative to the plain CRNN implementation.
As is typically found, some finer details at smaller scales remain unresolved that may be improved upon with further hyperparameter tuning and new modules.
Nonetheless, the improvement upon the baseline is substantial and our model shows promise for improving cardiac MRI reconstruction.

\section*{Acknowledgements}

This work was supported in part by National Institutes of Health (NIH) grant 7R01HL148788-03. C. Jordan, Y. Du and Y. Xue thank additional financial support from the School of Engineering, the University of Edinburgh. 
S.A.\ Tsaftaris also acknowledges the support of Canon Medical and the Royal Academy of Engineering and the Research Chairs and Senior Research Fellowships scheme (grant RCSRF1819\textbackslash 8\textbackslash 25).
The authors would like to thank Dr. Chen and K. Vilouras for inspirational discussions and assistance.

\newpage

\bibliographystyle{splncs04}
\bibliography{References}

\end{document}